\documentclass[twocolumn,prl,aps,showpacs]{revtex4}
\usepackage{graphicx}
\usepackage{amsmath}
\usepackage{sistyle}

\begin{document}

\title{High efficiency coupling of photon pairs in practice}

\author{T. Guerreiro, A. Martin, B. Sanguinetti, N. Bruno, H. Zbinden and~R.~T.~Thew}
\address{
Group of Applied Physics, University of Geneva, Switzerland}

\begin{abstract}
Multi-photon and quantum communication experiments such as loophole-free Bell tests and device independent quantum key distribution require entangled photon sources which display high coupling efficiency.
In this paper we put forward a simple quantum theoretical model which allows the experimenter to design a source with high pair coupling efficiency. In particular we apply this approach to a situation where high coupling has not been previously obtained: we demonstrate a symmetric coupling efficiency of more than 80\% in a highly frequency non-degenerate configuration. Furthermore, we demonstrate this technique in a broad range of configurations, \emph{i.e.} in continuous wave and pulsed pump regimes, and for different nonlinear crystals. 
\end{abstract}


\maketitle

\section{Introduction}

The efficient generation of photon pairs is fundamental to multi-photon experiments and the scaling of quantum networks. In particular, distributing quantum correlations through the use of photons is a promising resource for quantum cryptography technologies such as device independent quantum key distribution (DI-QKD)~\cite{Acin2006}. Thus, efficiently producing, distributing and detecting photon pairs from spontaneous parametric down conversion (SPDC)~\cite{Ljunggren2005, Bennink2010} is a major objective for current research, as an inefficient transmission or detection process opens loopholes in the verification of quantum correlations~\cite{Pomarico2011}.

The problem of detection has recently undergone significant progress, and single photon detectors with near unit efficiency have been demonstrated both for visible and telecommunication wavelengths~\cite{Lita2008,Miller2011,Fukuda2011}. The problem is therefore shifted towards the transmission of the photons. For long distance communication, single mode telecommunication fibers and wavelengths can be used~\cite{Takesue2010}. The single mode fibers have the additional advantage of guaranteeing the produced photon pairs to be in a single (spatial) mode, thus increasing the correlation in their detection. Residual loss can be compensated by heralding the presence of a photon with protocols such as qubit amplification~\cite{Bruno2013} or sum frequency generation \cite{Guerreiro2013}. 
This leaves the problem of efficiently coupling the photons into single mode optical fibers as the last problem to be solved.

To efficiently couple SPDC photon pairs into single mode optical fibers, they must be spatially correlated in such as way that if either photon is coupled in one fiber, the other is coupled in the other fiber. This is equivalent to maximizing the number of coincidences $C$ with respect to the number of single counts $R_\text{s}$ and $R_\text{i}$ for the signal and idler photons, respectively. This is usually called the symmetric heralding efficiency $\mu_{si}=C/ \sqrt{R_\text{s}\,R_\text{i}} $~\cite{Bennink2010}, and is the geometric average of the single heralding efficiencies defined as $ \mu_{s(i)}=C/ R_\text{s(i)} $.

Recently, general theoretical analyses have been put forward to solve this problem~\cite{Ljunggren2005, Bennink2010}. In these proposals it has been shown that there exists pump (p), signal (s) and idler (i) focusing parameters for which both the heralding efficiency into single mode optical fibers and the generation rates are high. In addition, high heralding efficiency has been demonstrated experimentally~\cite{Wittmann2012, Giustina2013, Pereira2013, Christensen2013}.
These articles, however, do not provide a practical methodology to obtain high heralding efficiencies and the general theoretical treatments presented in~\cite{Ljunggren2005, Bennink2010} still require considerable adaptation on an experimental level.
In this article we show how $\mu_{si}$ can be maximized. The method which we describe should give the reader a good intuition and understanding of how to achieve high heralding efficiency in practice, with a variety of crystals and in different experimental configurations. As will become clear later on, this method relies on the entanglement of the generated photons, and is thus quantum mechanical.

In what follows, we explain in simple terms the intuition behind obtaining high coupling through this method, and show how phasematching calculations are used to determine the pump, signal and idler focusing parameters that maximize heralding efficiency. We then follow describing the experimental implementation of these ideas in a highly non degenerate SPDC setup (for which high $\mu_{si}$ has not been demonstrated in the literature) based on potassium niobate ($\rm KNbO_3$, type-I angular phasematching) and periodically poled lithium niobate (PPLN, type-0 quasi-phasematching). We do this in two different pump regimes, continuous wave (CW) and pulsed, and demonstrate $ \mu_{s(i)} > 80\% $ and $ \mu_{si} \sim 80\% $.

\section{Intuition}

In this section we explain, in terms of a simple one-dimensional model, the intuition behind the technique used here to achieve high heralding efficiency. 

Let's consider the SPDC process, where a pump photon ($ p $) interacts with a non-linear medium to decay into a pair of signal ($ s $) and idler ($ i $) photons. If the pump has a well defined wavevector (i.e. it can be approximated as a plane wave), and the energy of the photons is uncorrelated with their wavevector directions, the resulting state can be written as a correlated two-photon wavefunction in the wavevector basis \cite{Walborn2010}:
\begin{eqnarray}
\vert \Psi \rangle_{s,i} = \int dk \vert k \rangle_{s} \vert k_{p} - k \rangle_{i}.
\label{state}
\end{eqnarray}
Now we want to ``project'' the idler photon in the state corresponding to the fundamental mode of an optical fiber, denoted by $ \vert \phi \rangle_{i} $. The representation of this state in the position basis is
\begin{eqnarray}
\phi_{i}(x) = \langle x \vert \phi \rangle_{i} = \dfrac{1}{(2\pi\sigma^{2})^{1/4}} e^{-x^{2}/4\sigma^{2}},
\label{project_position}
\end{eqnarray}
where $ \sigma $ is the width of the Gaussian packet. Writing (\ref{project_position}) in the momentum basis
\begin{eqnarray}
\phi_{i}(k) = \langle k \vert \phi \rangle_{i} = \left( \dfrac{2}{\pi} \right)^{1/4} \sigma^{1/2} e^{-k^{2}\sigma^{2}},
\label{project_momentum}
\end{eqnarray}
we can calculate the resulting state of the signal photon $ \Psi_{s}(x) = \langle x \vert _{i}\langle \phi \vert \Psi \rangle_{s,i} $ after the idler photon has been projected onto the state $ \vert \phi \rangle_{i} $:
\begin{eqnarray}
 \Psi_{s}(x)& = & \int dk \langle x \vert k \rangle_{s} \langle \phi \vert k_{p} - k \rangle_{i} \\ 
& = & \dfrac{e^{i k_{p} x}}{(2\pi\sigma^{2})^{1/4}} e^{-x^{2}/4\sigma^{2}} \label{s photon}
\end{eqnarray}
It is possible to see from equation (\ref{s photon}) that every time photon $ i $ is measured to be in the Gaussian state (\ref{project_position}), photon $ s $ must also be found in a Gaussian state of the same form up to a global phase. This means that once photon $ i $ is detected to be in the fundamental mode of an optical fiber, photon $ s $ will be heralded in the fundamental mode of a fiber, leading to a high signal-idler coincidence and a high heralding efficiency. Of course, the effect is symmetric under exchange $ s \leftrightarrow i $, so it does not matter if the idler photon is used to herald the signal photon or the opposite.

It is the correlations in the state (\ref{state}), generated by a plane wave pump which leads to this high probability of heralding a photon in the mode of an optical fiber. If such a state cannot be generated, the above method will not work. Of particular interest is the situation in which the wavevectors of the signal and idler photons are correlated to their respective wavelengths. If this is the case, the detection of the idler photon in a state $ \vert \phi \rangle_{i} $ will project the signal photon onto an entangled state between wavevector and wavelength; once the information about the wavelength is traced out, the signal photon will be left in a statistical mixture of different wavevectors, which will not couple well into the fundamental mode of an optical fiber. 

With this we see that two points are crucial to efficiently herald a photon in the mode of an optical fiber through the method described above: first, it is necessary to pump the non-linear medium with a plane-wave (collimated) pump. Second, it is fundamentally important to have photons in a wavelength-wavevector separable state, thus working in a collection regime where these spectral-spatial correlations are reduced as much as possible. In the following sections we address these two points.

\section{Choosing pump focusing and collection parameters}

To address the above-mentioned issues we considered the phasematching of our crystal (we used a custom-made program but there are programs which can be found online, for example \cite{phasematch}). In the first instance we use a specific nonlinear crystal and pump regime, though the approach here is general and can be applied to a vast number of configurations. The success of the methods described depend only on the crystal length, refractive index, pump focusing and collection optics of the setup.

As an example, we consider the phase-matching for the interaction:
\[\SI{532}{nm} \rightarrow \SI{810}{nm}+\SI{1550}{nm}\] in a \SI{10}{mm} long $\rm KNbO_3$ crystal pumped in pulsed regime (\SI{8}{ps}). In what follows we determine the pump focusing and the collection waists for this example, but we note that the same results for pump focusing and collection are obtained if we considered as an example the case of CW pump regime.

\subsection{Pump focusing}
To maximise the correlations between $s$ and $i$, the pump field should be close to a plane wave. In this section we examine to what extent the pump wavefront should be plane for the photon coupling to be good.

First, we define the focusing parameter, which tells how much a beam is focused inside a non-linear crystal with respect to the crystal length. This definition follows the one presented in \cite{Bennink2010}. The focusing parameter denoted $ \xi $ is simply the half length of the crystal $ L/2 $ divided by the Rayleigh range $ z_{R} $ of the beam
\begin{eqnarray}
\xi = \dfrac{L}{2 \cdot z_{R}} = \dfrac{\lambda L}{2\pi w_{0}^{2}}
\end{eqnarray}
where $ \lambda $ and $ w_{0} $ correspond to the wavelength and waist at the center of the crystal, respectively. The closer $ \xi $ is to zero, the more the beam can be approximated by a plane wave. On the other hand, if $ \xi \gg 0 $ the beam is tightly focused and localized at a point in space.

\begin{figure}[h!]
\includegraphics[width=\columnwidth]{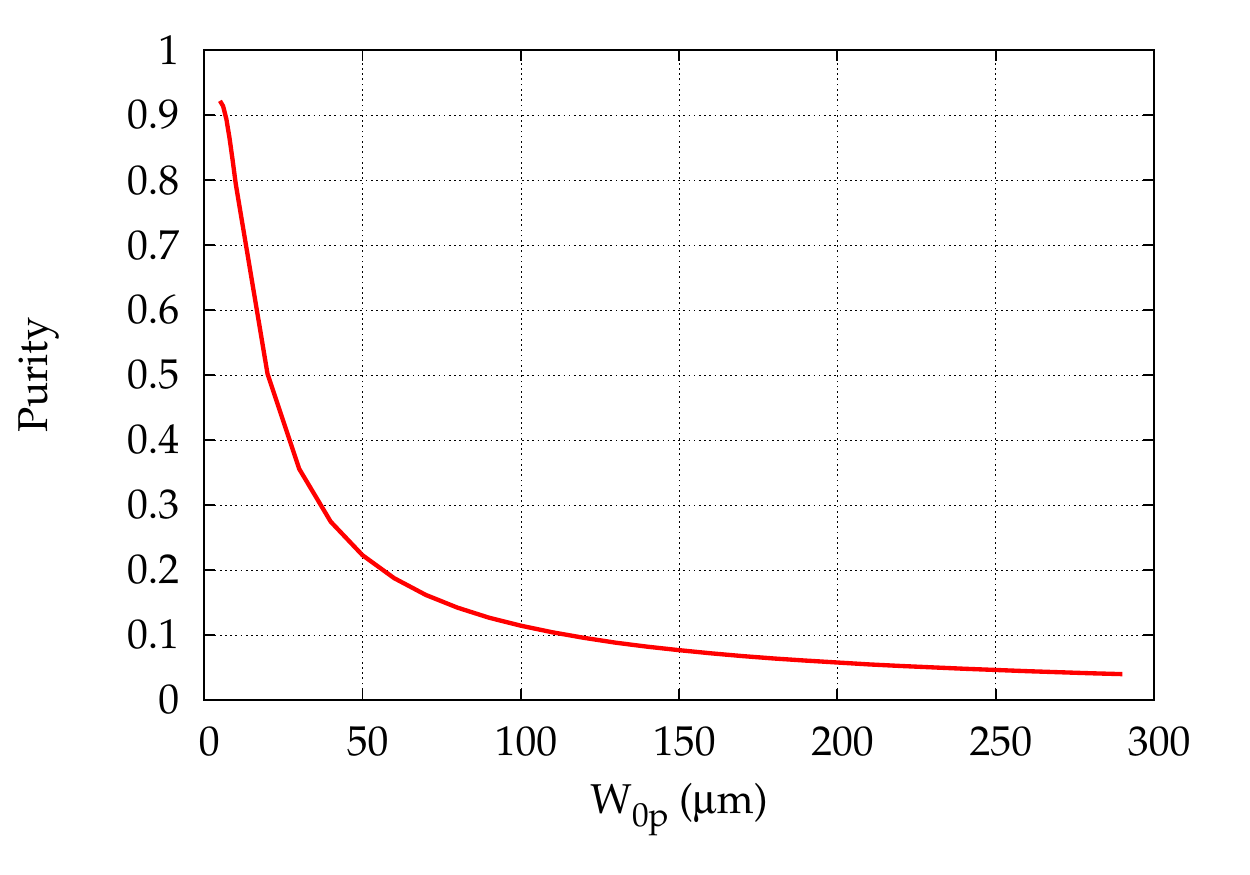}
\centering
\caption{Purity of the joint signal-idler angular correlation function as a function of the pump waist. Correlation is higher with large pump waists: we wish to operate as close to this regime as possible.}
\label{angular_purity}
\end{figure}

To achieve a high wavevector signal-idler correlation, and consequently a high heralding efficiency, it is necessary to operate in a regime where the pump focusing parameter is close to zero. \figurename{~\ref{angular_purity}} shows the degree of separability, or purity (analogous to the one defined in \cite{Mosley2008}), of the two photon joint angular correlation function, as a function of the pump waist. We can see that the larger the pump waist (the lower $ \xi_{p} $), the lower the purity and consequently the higher the correlation in wavevectors of signal and idler. In the limit that the waist of the pump is infinite (plane wave) we have zero purity and approach a perfectly entangled state like the one in equation (\ref{state}). 

There are, however, limits on the pump focusing parameter that one should keep in mind. First, the lower the focusing of the pump, the smaller the generation rates will be. Second, special attention must be taken about the cross section of the crystal: if the pump focusing is too low, the beam might hit the edges of the crystal. These two observations set a practical upper limit on the pump focusing. A third point, which sets a lower limit on the pump focusing, is the uncertainty in the wavevector: the pump wavevector uncertainty cannot be higher than the wavevector uncertainty in the collection modes of the signal and idler photons. This wavevector uncertainty corresponds to the angular spread in the case of a Gaussian beam.

In our setup we would like to optimise only the heralding efficiency. For this reason, it suffices to set the pump angular spread much smaller that the collection modes angular spread, keeping in mind the restrictions given by the cross section of the nonlinear crystal. Taking these points into consideration, a good value for pump focusing in our setup (for both PPLN and $ \rm KNbO_{3} $) was $ \xi_{p} = 0.02 $.

\subsection{Signal and Idler collection}

As mentioned above, the second point is to operate in a collection regime where the wavelength of the signal (idler) photon is uncorrelated from its wavevector direction. To verify this we considered the phase-matching of our crystal and calculated the correlations between angle of emission and the photon's wavelengths. As we will show below, these calculations will determine the size of the image of the mode of the optical fiber at the center of the nonlinear crystal and consequently all the lenses of the setup. Since we chose $ \xi_{p} = 0.02 $ this implies a waist for the pump of \SI{200}{\mu m}, which yields an angular spread of $ \Delta \Theta_{p} \sim 0.0017 $ rad.

\begin{figure}[h!]
\includegraphics[width=\columnwidth]{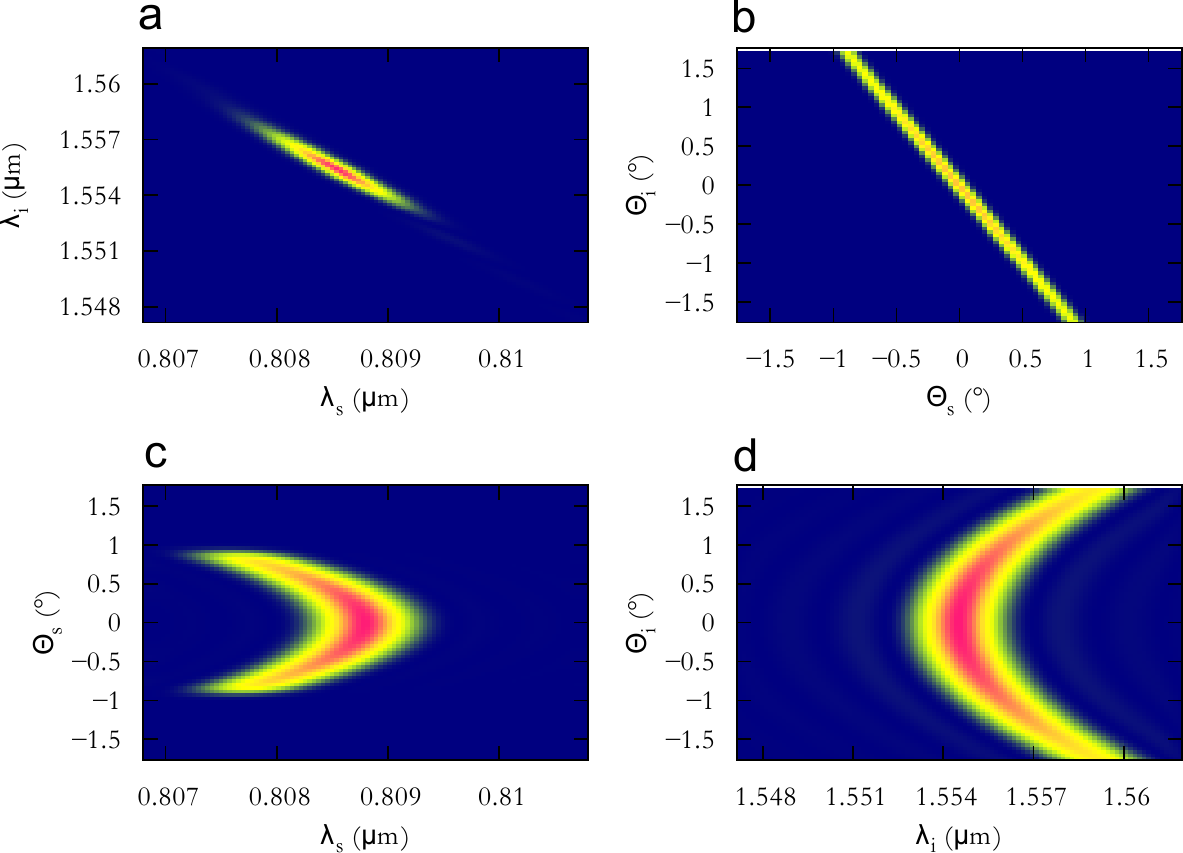}
\centering
\caption{\textbf{a}, $\rm KNbO_3$ joint spectral function of the two photons in pulsed regime. \textbf{b}, Joint angle of emission function. \textbf{c}, Correlation (spectral-spatial) function between the angle of emission and wavelength of signal photon. For a certain range of emission angles corresponding to $ 0.2^{\circ} $ the emission angle is independent of the wavelength. \textbf{d}, Correlation function between the angle of emission and wavelength of idler photon. All angles are external to the crystal.}
\label{spectral_spatial}
\end{figure}

In \figurename{~\ref{spectral_spatial}} it is possible to see the complete characterization of the $ \rm KNbO_{3} $ crystal pumped in pulsed regime. \figurename{~\ref{spectral_spatial}\textbf{a}} shows the joint spectral function while \ref{spectral_spatial}\textbf{b} shows the angular correlation intensity function. \figurename{~\ref{spectral_spatial}\textbf{c}} and \ref{spectral_spatial}\textbf{d} present the spectral-spatial correlations for signal and idler photons, respectively. These functions are all we need to determine the collection waists of the setup. All the angles involved are external to the nonlinear crystal.

To choose a collection waist for the signal photon, we look at \figurename{~\ref{spectral_spatial}\textbf{c}}, where it is possible to see that for a certain range of emission angles, approximately from $ -0.1^{\circ} $ to $ 0.1^{\circ} $ ($ \Delta \Theta_{s} \sim 0.2^{\circ} \sim 0.0035 $ rad) the angle of emission is independent of the wavelength, \textit{i.e.} the photon is pure in this basis. This determines the collection waist for the signal photon to be approximately
\begin{equation}
w_{0,s} \simeq \dfrac{2 \lambda_{s}}{\pi \cdot \rm \Delta \Theta_{s}} \simeq \SI{145}{\mu m}
\end{equation}

\begin{figure}[h!]
\includegraphics[width=\columnwidth]{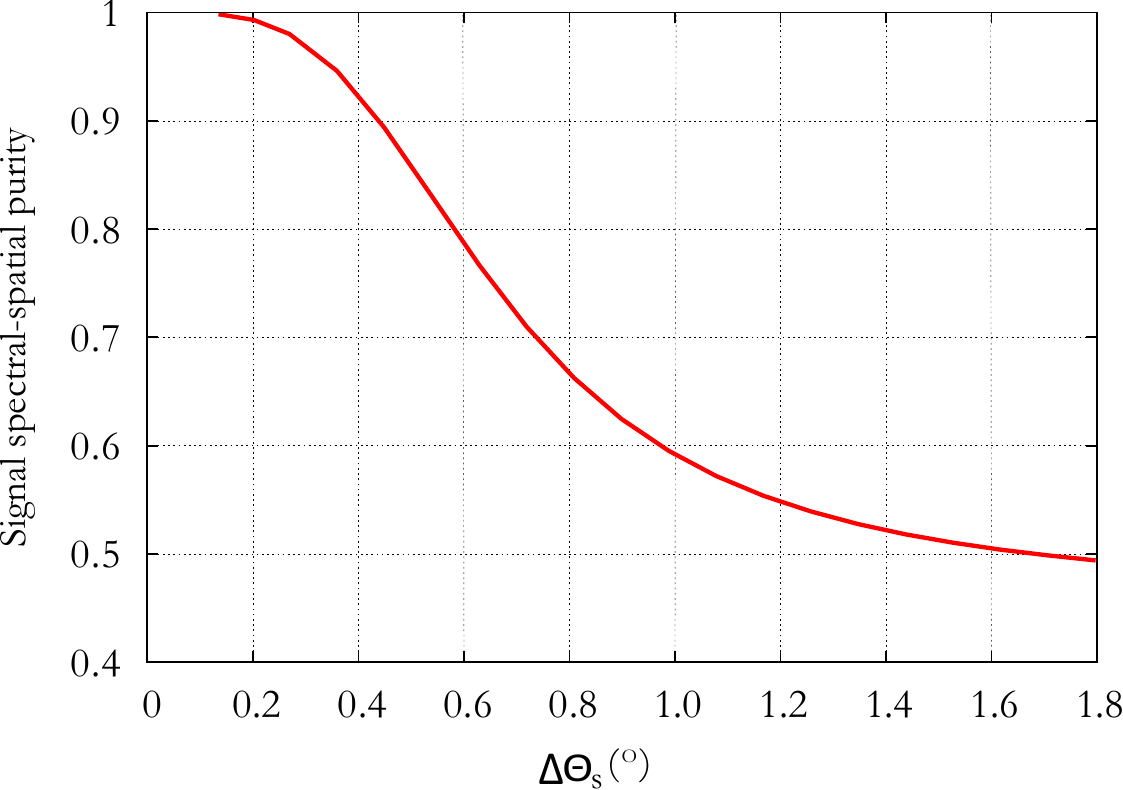}
\centering
\caption{Purity of the spectral-spatial wavefunction as a function of the signal angular collection. The purity starts to drop after the signal angular collection becomes $ \sim 0.3^{\circ} $, which justifies the choice $ \Delta \Theta_{s} = 0.2^{\circ} $}
\label{purity}
\end{figure}

This choice can be readily justified if we look at \figurename{~\ref{purity}}, where the spectral-spatial ``purity'' of the state is shown as a function of $ \Delta \Theta_{s} $. As $ \Delta \Theta_{s} $ increases, the spectral-spatial purity (and consequently the heralding efficiency) decreases and stabilizes at a value around $ 50 \% $. To determine the corresponding collection waist for the idler we look at the joint signal idler angle correlation function in \figurename{~\ref{spectral_spatial} \textbf{b}}, where we can see then that the corresponding $ \Delta \Theta_{i} $ for the idler is approximately twice the one for the signal $ \Delta \Theta_{i}~\simeq~2~\cdot~\Delta~\Theta_{s}~\simeq~0.4^{\circ} $ ($ = 0.007$ rad), giving an idler collection waist of
\begin{equation}
w_{0,i} \simeq \dfrac{2 \lambda_{i}}{\pi \cdot \rm \Delta \Theta_{i}} \simeq \SI{140}{\mu m}
\end{equation}

\begin{figure}[h!]
\includegraphics[width=\columnwidth]{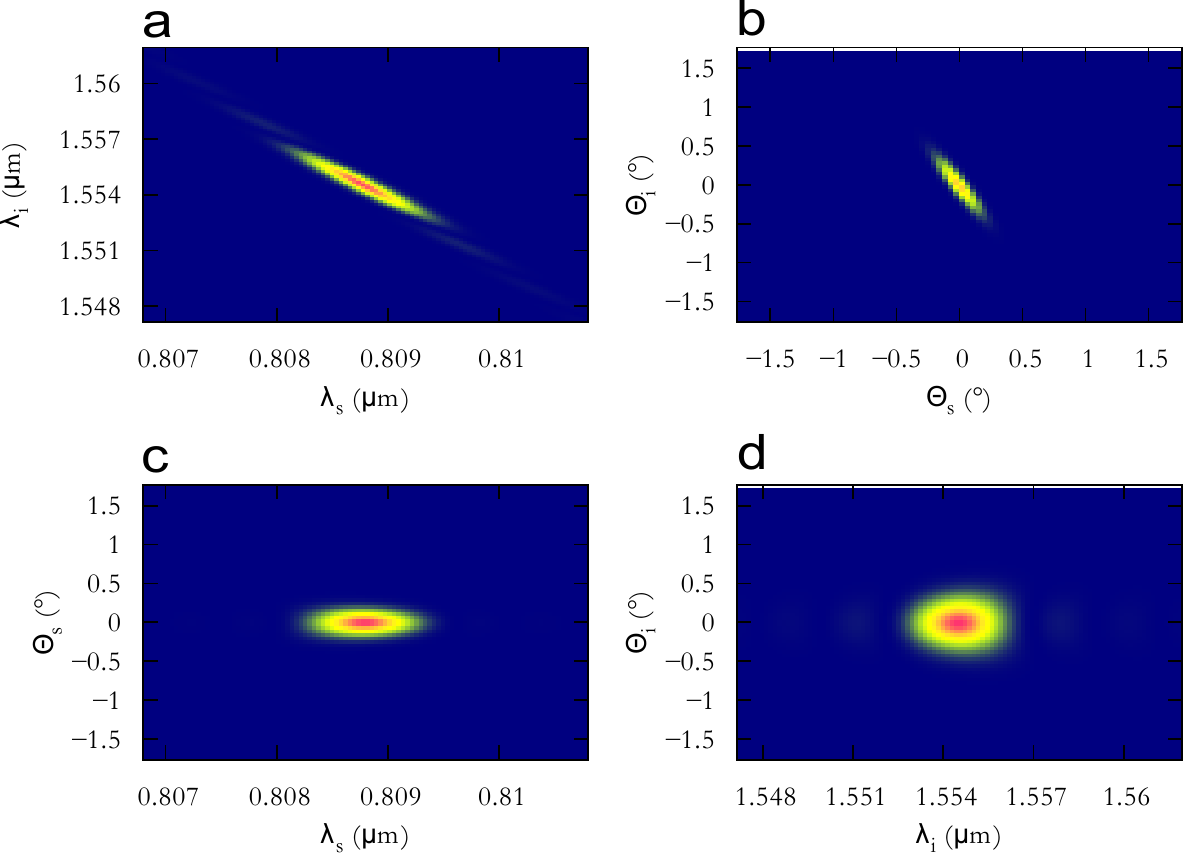}
\centering
\caption{Phasematching functions after coupling into single mode optical fibers, collection waists $ w_{0,s} \simeq \SI{145}{\micro m} $ for signal and $ w_{0,s} \simeq \SI{140}{\micro m} $ for idler.}
\label{spatial}
\end{figure}

\figurename{~\ref{spatial}} shows the various phasematching functions after imaging these collection waists in single mode optical fibers. Notice that the joint spectral function, depicted in \figurename{~\ref{spatial}\textbf{a}} remains unchanged after coupling into the fiber. On the other hand, the emission angle of signal and idler photons become uncorrelated from the spectrum, as can be seen in \figurename{~\ref{spatial}\textbf{c}} and \ref{spatial}\textbf{d}. 

If we consider the focusing parameters for the pump ($ w_{0,p} \simeq $~\SI{200}{\mu m}), signal ($ w_{0,s} \simeq \SI{145}{\mu m} $) and idler ($ w_{0,i}~\simeq~\SI{140}{\mu m} $) with a crystal length of \SI{10}{mm}, we find $ \xi_{532} = 0.02 $, $ \xi_{810} = 0.06 $ and $ \xi_{1550}~=~0.13 $, respectively. This shows that our considerations are outside any of the optimal values predicted by \cite{Ljunggren2005, Bennink2010}, since here we only optimize the heralding efficiency and not the generation rates.

\section{Experiment}

\begin{figure}[h!]
\includegraphics[width=1\columnwidth]{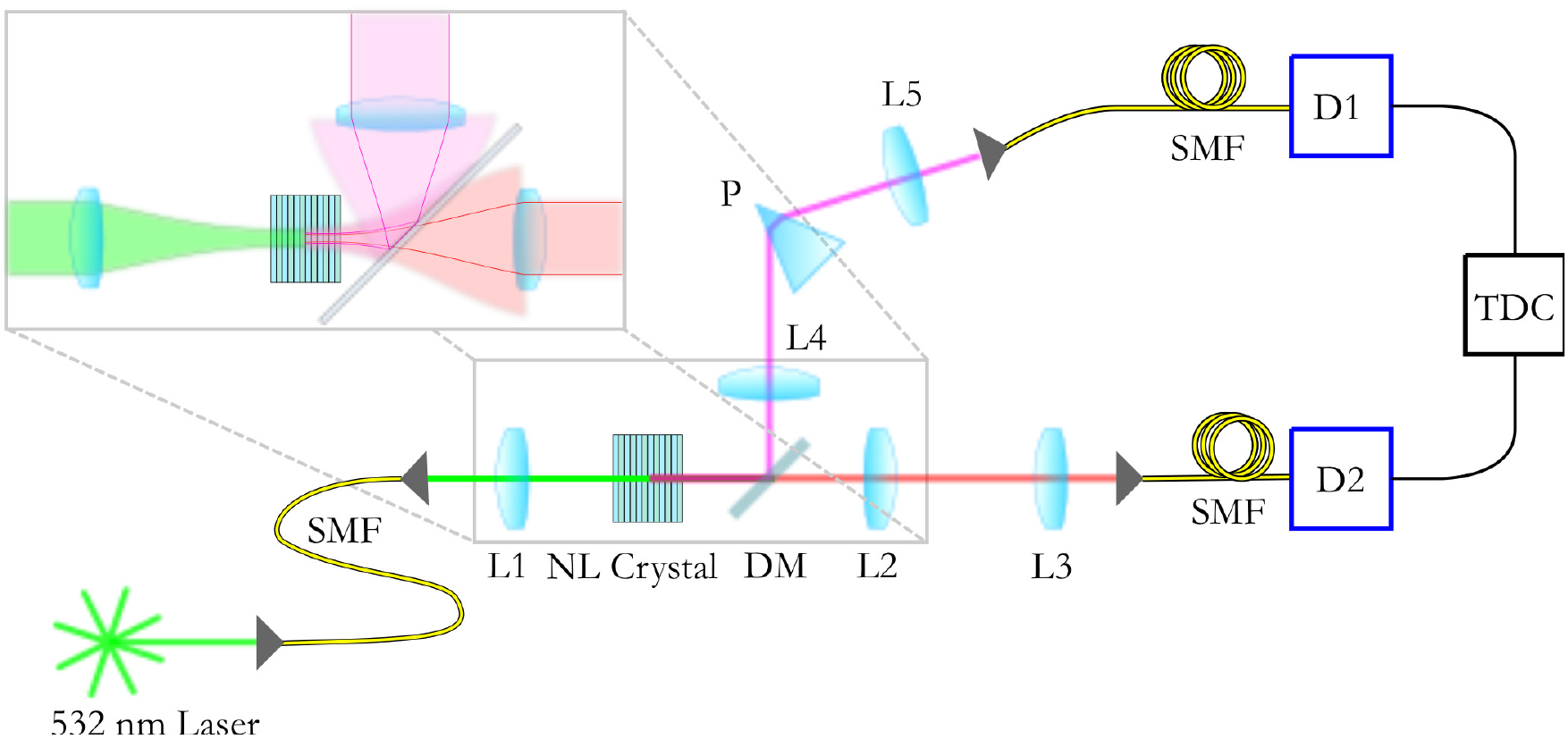}
\caption{ \label{setup} Experimental setup. A \SI{532}{nm} laser passes through a single mode optical fiber at that wavelength, and is used to pump a nonlinear crystal. The remaining pump light is removed by a high pass filter, and the signal and idler photons are split by dichroic mirror (DM). The transmitted photons (\SI{1550}{nm}) are collimated by the lens L2 and focused on the fiber by L3. The reflected photons (\SI{810}{nm}) are collimated by L4 and focused inside the fiber by L5. The zoom shows the output of the nonlinear crystal, where emission happens through a wide range of angles. Only a small fraction of the emitted light, as determined above, is imaged (collected) into the optical fibers.}
\end{figure}

A schematic of the experimental setup is shown in \figurename{~\ref{setup}}. A laser at \SI{532}{nm} passes through a single mode optical fiber to ensure a perfect Gaussian profile. The output of the fiber is directed to an aspheric lens of focal $f=\SI{15.29}{mm}$ (L1), to produce a \SI{200}{\mu m} waist at the center of a \SI{10}{mm} long nonlinear crystal, either a PPLN with cross section \SI{0.5}{mm}$ \times $\SI{0.5}{mm} or a $ \rm KNbO_{3} $, with cross section \SI{5}{mm}$ \times $\SI{5}{mm}. The waist of the pump at the position of the center of the crystal is verified with a CCD camera. 

The crystal produces pairs of photons at \SI{1550}{nm} and \SI{810}{nm}, which are collected from a waist of $ \sim $\SI{140}{\mu m} and $ \sim $\SI{145}{\mu m} respectively, as calculated above. To image these collection waists on the optical fibers, we use two lenses on each path. On the \SI{1550}{nm} path, an $ f~=~\SI{150}{mm}$ lens (L2) is used to collimate the beam, and an $ f=\SI{7.5}{mm}$ lens (L3) is used to focus the beam on the optical fiber. These lenses were selected taking into account the mode field diameter of the telecom single mode fiber (\SI{5.1}{\mu m}) and were verified to produce the desired waist at the crystal position during the alignment, using a CCD camera. The \SI{810}{nm} path has a similar setup, with an $f=\SI{150}{mm}$ lens (L4) to collimate and an $ f=\SI{3.1}{mm}$ lens (L5) to focus. Again, lenses were chosen according to single mode fiber diameter (\SI{2.8}{\mu m}) and the produced waist was verified with a CCD. After the crystal, the remaining pump light is filtered by a high-pass filter and a prism (P) on the \SI{810}{nm} path. 

When coupled into fibers, the photons are then directed to detectors, connected to a time-to-digital converter (TDC). For the \SI{810}{nm} photon we use a free running Si-APD (D1), with efficiency of $ (48.0 \pm 2.5 ) \% $ at \SI{810}{nm}. On the \SI{1550}{nm} side we use an InGaAs detector (D2), with efficiency of $ (24 \pm 2) \% $. The heralding efficiency $ \mu_{s(i)} $ for the signal (idler) is defined to be the ratio between coincidences $ C $ to the number of singles (detector noise subtracted) in the idler (signal) detector $ S_{i(s)} $ corrected by the detection efficiency $ \eta_{s(i)} $. We are interested in characterizing the coupling into the optical fibers. For this reason we also correct for the transmissions in the signal (idler) path $ t_{s(i)} $: 

\begin{eqnarray}
\mu_{s(i)} = \dfrac{C}{R_{s(i)}} = \dfrac{C}{S_{i(s)} \cdot \eta_{s(i)} \cdot t_{s(i)}}
\end{eqnarray}
The measured values for transmissions in the setup are $ t_{1550}~=~(87.0 \pm 0.2) \% $ and $ t_{810}~=~(78.0~\pm~0.2)~\% $. These values are limited mainly by the transmission of the high pass filter used to remove the pump.

If we consider the $\rm KNbO_3$ crystal pumped by a CW laser at \SI{5}{mW} the rate of singles at \SI{810}{nm} was \SI{39.0}{KHz}, while the coincidence rate was around \SI{7.0}{KHz}. Integrating for \SI{30}{s} a heralding efficiency of $ \mu_{1550} = (86 \pm 7) \% $ was measured. The error on this value is dominated by the systematic error in the measurement of the detector efficiency. If we reverse the detection roles, heralding the \SI{810}{nm} photons with those at \SI{1550}{nm}, the rate of singles was \SI{3.2}{KHz}, while the coincidence rate was around \SI{0.9}{KHz}. Integrating also for \SI{30}{s} a heralding efficiency of $ \mu_{810} = (75 \pm 5) \% $ was measured. This yields a symmetric heralding efficiency of $ \mu_{si}~=~\sqrt{\mu_{s} \cdot \mu_{i}}~\sim~80\% $. The same heralding efficiencies are obtained in both CW and pulsed regimes. 

\section{Summary of the method}

We can summarize the technique to achieve high heralding efficiency in the following simple steps. 

\begin{itemize}
\item[1)] The wavevector uncertainty for the pump must be smaller than the collection modes' wavevector uncertainty. This means the angular spread for the pump must be smaller than the angular spread of the signal and idler collection modes, a condition which can be achieved by setting the pump focusing parameter close to zero. In our setup we have about $\xi_{p}\sim~0.02 $. 
\item[2)] Calculate the signal spectral-spatial correlation function for the crystal used (shown here in \figurename{~\ref{spectral_spatial}\textbf{c}}). This will determine the signal angular collection and consequently the signal collection waist.
\item[3)] Calculate the signal-idler joint angle correlation function. Given the signal angular collection, this function will determine the idler angular collection and consequently the idler collection waist.
\item[4)] Purity is highest towards the centre of the beam. For this reason, in the presence of wavelength dependent optical elements, it is very important to align with the exact wavelengths which will be collected into the fiber.
\end{itemize}

This method was described using a \SI{10}{mm} $\rm KNbO_{3}$ crystal (type-I, frequency non-degenerate angular phasematching) as an example, pumped in CW and pulsed (\SI{8}{ps}) regimes. To test the universality of this approach we also verified the technique using a \SI{10}{mm} PPLN (type~-~0, frequency non-degenerate quasi-phasematching) pumped in pulsed (\SI{8}{ps}) regime. In addition to that, we have also employed the same method in a completely different setup, using a pulsed (\SI{2}{ps}) laser at \SI{780}{nm} to pump a \SI{30}{mm} PPKTP crystal with type-II degenerate quasi-phasematching, generating photons at \SI{1560}{nm}. The design of these sources using this technique all yielded experimental results for $ \eta_{si} \sim 80 \% $. We summarize the different tested configurations in Table \ref{table}.
\begin{table*}
\centering
\begin{center}
\begin{tabular}{l*{6}{c}r}
NL crystal    & Length       & Pump & Phasematching & Type & $ \eta_{si} $  \\
\hline
\hline
$ \rm KNbO_{3} $ & \SI{10}{mm} & CW & $\SI{532}{nm} \rightarrow \SI{810}{nm}+\SI{1550}{nm}$ & I & $ \sim $80 \%   \\
$ \rm KNbO_{3} $ & \SI{10}{mm} & Pulsed, \SI{8}{ps} & $\SI{532}{nm} \rightarrow \SI{810}{nm}+\SI{1550}{nm}$ & I & $ \sim $80 \%   \\
PPLN & \SI{10}{mm} & Pulsed, \SI{8}{ps} & $\SI{532}{nm} \rightarrow \SI{810}{nm}+\SI{1550}{nm}$ & 0 & $ \sim $80 \%  \\
PPKTP & \SI{30}{mm} & Pulsed, \SI{2}{ps} & $\SI{780}{nm} \rightarrow \SI{1560}{nm}+\SI{1560}{nm}$ & II & $ \sim $80 \%  \\
\end{tabular}
\caption{ \label{table} Different setups where the proposed method was employed. The wide range of configurations proves the universality of the technique.}
\end{center}
\end{table*}

Notice that for the PPKTP, the length of \SI{30}{mm} imposes a different pump waist from the one used in the examples described in the last sections, and the simulations shown in figure~\ref{spectral_spatial} yield different results, thus implying other collection conditions. This provides strong experimental evidence for the universality of this method, showing that the above four steps can be employed in a wide range of configurations to obtain high heralding efficiencies.

\section{Conclusion}

In this paper we report a method to design a high heralding efficiency photon pair source for bulk crystals. This technique exploits the purity of photons in the wavelength-wavevector basis and the signal-idler correlations in wavevector. 

A theoretical and experimental study of different crystals, for both degenerate and non-degenerate photon pairs as well as pulsed and CW regimes shows that the method works in a broad range of configurations. The main parameters are the pump focusing, the length and the refractive index of the crystal. Given these parameters we show how to determine the signal and idler collection waists and consequently all the lenses of the setup. This work has important implications for anyone designing a photon pair source based on SPDC bulk crystals, and can be applied to a wide range of experiments.

\textbf{Acknowledgments}: We are thankful to Andre Stefanov for helpful discussions. This work was supported by the Swiss NCCR QSIT project.


\end{document}